# Designing Open Source Computer Models for Physics by Inquiry using Easy Java Simulation

Loo Kang WEE[1], Sze Yee LYE[1]

[1]Ministry of Education, Education Technology Division, Singapore

wee_loo_kang@moe.gov.sg, lye_sze_yee@moe.gov.sg

http://weelookang.blogspot.sg, http://iwant2study.org/easyjava

Abstract: The Open Source Physics community (http://www.compadre.org/osp/) has created hundreds of physics computer models (Wolfgang Christian, Esquembre, & Barbato, 2011; F. K. Hwang & Esquembre, 2003) which are mathematical computation representations of real-life Physics phenomenon. Since the source codes are available and can be modified for redistribution licensed Creative Commons Attribution or other compatibile copyrights like GNU General Public License (GPL), educators can customize (Wee & Mak, 2009) these models for more targeted productive (Wee, 2012) activities for their classroom teaching and redistribute them to benefit all humankind.

In this interactive event, we will share the basics of using the free authoring toolkit called "Easy Java Simulation" (W. Christian, Esquembre, & Mason, 2010; Esquembre, 2010) (http://www.um.es/fem/EjsWiki/pmwiki.php) so that participants can modify the open source computer models for their own learning and teaching needs. These computer models has the potential to provide the experience and context, essential for deepening student's conceptual understanding of Physics through student centred guided inquiry approach (Eick, Meadows, & Balkcom, 2005; Jackson, Dukerich, & Hestenes, 2008; McDermott, Shaffer, & Rosenquist, 1995; Wee, Lee, & Goh, 2011).

.

Keyword: easy java simulation, active learning, education, teacher professional development, e-learning, applet, design, open source, physics
PACS: 01.50.H-, 07.05.Tp, 01.50.Lc, 83.10.Rs

## I. INTRODUCTION

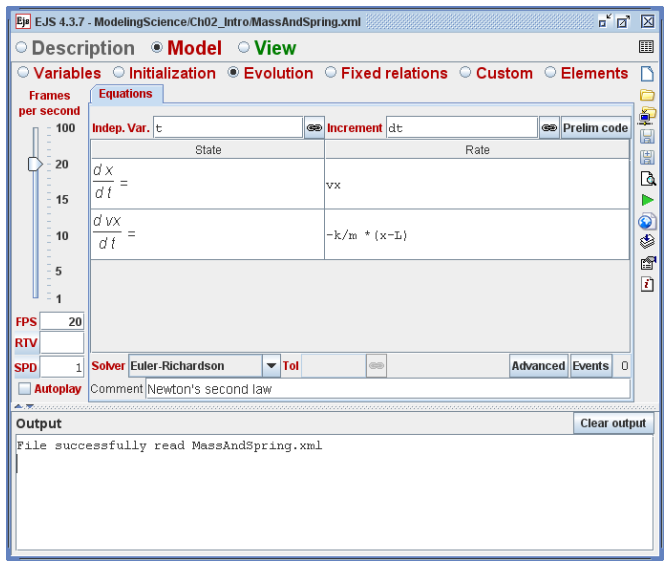

Figure 1. EJS authoring toolkit version 4.3.7 showing the Evolution Page with the Ordinary Differential Equations to model a simple spring mass system.

The goal of the Open Source Physics (OSP) project is to make a large number of simulations together with source code available for education using the GNU General Public License (GPL) open-source model. OSP provides both high-level modeling tools and a lower-level computational physics library to create computer simulations through the use of a consistent object-oriented framework (Wolfgang Christian & Esquembre, 2012). The Java-based OSP library defines objects to build interactive user interfaces, draw 2D and 3D objects, numerically solve ordinary differential equations using different algorithms, and represent data using tables and graphs.

This workshop provides a hands-on beginners' introduction to Open Source Physics (OSP) and Easy Java Simulations (EJS) to model physical systems such as those provided by the interactive event leaders.

Participants will study and explore, step by step, important computational example of the spring mass system model (Wolfgang Christian & Esquembre, 2008a, 2008b), to learn how they have been implemented, and then modify these example to add new capabilities. Assistance will be provided during the sessions. There after, the organisers hope to get participants to identify and self direct (Tan, Shanti, Tan, & Cheah, 2011) their own personally motivating computer model from the OSP and NTNU digital libraries and work collaboratively (Chai, Lim, So, & Cheah, 2011) with the OSP community to create customized or new models.

During the workshop we will discuss the general pedagogical (Wee, Chew, Goh, Tan, & Lee, 2012) and technical issues in the design of interactive computer-based tutorials as well as how existing models can be adapted to Singapore school context. All workshop material will be made available through our blogs http://weelookang.blogspot.sg and http://iwant2study.org/easyjava .

## II. DESCRIPTION OF FORMAT OF INTERACTIVE EVENT

The interactive event consists of the following segments:

- Introduction to Open Source Physics
- Sharing of computer models customized by the interactive event leaders
- Introduction of Easy Java Simulation

- Hands-on-with Easy Java Simulation with spring mass system model (Wolfgang Christian & Esquembre, 2008a, 2008b)
- Sharing of tips to easily adapt codes from other simulations into a new model
- Participants download a simulation that is personally motivating to them from OSP Digital Libraries, that they want to edit and improve on even after the workshop, in collaborative group of 2-3.
- Upload to NTNU Java Virtual Lab (F.-K. Hwang, 2010) (http://www.phy.ntnu.edu.tw/ntnujava/index.php?board=28.0)
- Closing discussions

III.   MATERIALS, HARDWARE, SOFTWARE, AND OTHER TECHNOLOGIES

- Easy Java Simulation (Esquembre, 2010) (EJS) tool kit
- Java Runtime
- Java 3D (for Java 3D models)

IV.   SCHEDULING NEED

- One shot (3 hour or half day)

V.   SUMMARY OF EQUIPMENT

- BYOL (Bring Your Own Laptop)
- Wireless access

VI.   CONCLUSION

This is a series of professional workshops that is conducted by the interactive event leaders in a funded project by NRF2011-EDU001-EL001 Java Simulation Design for Teaching and Learning, (MOE, 2012b) awarded by the National Research Foundation in collaboration with National Institute of Education, Singapore and the Ministry of Education (MOE), Singapore.

Our computer models are downloadable on NTNU Java Virtual Lab (F.-K. Hwang, 2010) creative commons attribution licensed and lesson packages on the ICT Connection edumall 2.0 portal under ICT-in-ACTION http://ictconnection.edumall.sg/cos/o.x?c=/ictconnection/ictlib&uid=200&ptid=711 (edumall 2.0 login required).

We hope other educators will share their own creative works with the world as demonstrated by the OSP community leaders like Francisco Esquembre, Fu-Kwun Hwang and Wolfgang Christian and the passionate (interactive event leaders included) community of the Open Source Physics Project.

ACKNOWLEDGEMENT

We wish to acknowledge the passionate contributions of Francisco Esquembre, Fu-Kwun Hwang and Wolfgang Christian for their ideas and insights in the co-creation of interactive simulation and curriculum materials.

This workshop research is made possible thanks to the eduLab project NRF2011-EDU001-EL001 Java Simulation Design for Teaching and Learning, (MOE, 2012b) awarded by the National Research Foundation in collaboration with

National Institute of Education, Singapore and the Ministry of Education (MOE), Singapore.

Lastly, we also thank MOE for the recognition of our research on the computer model lessons as a significant innovation in 2012 MOE Innergy (HQ) GOLD Awards (MOE, 2012a) by Educational Technology Division and Academy of Singapore Teachers.

## INTERACTIVE EVENT LEADERS (AUTHORS)

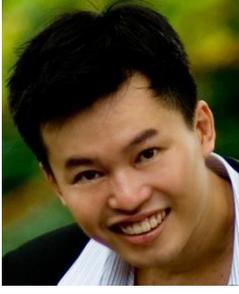
Loo Kang WEE is currently an educational technology specialist at the Ministry of Education, Singapore and a PhD candidate at the National Institute of Education, Singapore. He was a junior college physics lecturer and his research interest is in Open Source Physics tools like Easy Java Simulation for designing computer models and use of Tracker.

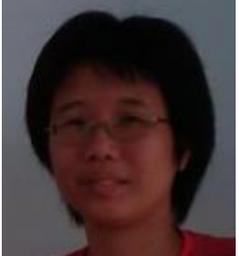
Sze Yee LYE is currently educational technology officer at the Ministry of Education, Singapore. She was a primary and secondary physics teacher and her research interest is in free tools for technology use in education.